\documentclass{iau}
\usepackage{graphicx}

\title
[IAU 317.~~New axes for the fundamental plane]
{New axes for the stellar mass \\ fundamental plane}

\author[Schechter]
{Paul L. Schechter$^1$}

\affiliation{$^1$MIT Kavli Institute Cambridge, MA 02139, USA \\ 
email: {\tt schech@mit.edu}\\
}

\pubyear{2015}
\volume{317}  
\pagerange{901--904}
\jname{The General Assembly of Galaxy Halos}
\editors{
M. Arnaboldi,
\ A.  Bragaglia,
\ M. Rejkuba, \& 
\ D. Romano, 
 eds.}
\begin{document}

\maketitle

\begin{abstract}
We argue that the stellar velocity dispersion observed in an elliptical galaxy
is a good proxy for the halo velocity dispersion.  As dark
matter halos are almost completely characterized by a single scale
parameter, the stellar velocity dispersion tells us the virial radius of the 
halo and the mass contained within.  This permits non-dimensionalizing of the
stellar mass and effective radius axes of the stellar mass fundamental 
plane by the virial radius and halo mass, respectively.


\end{abstract}

\firstsection 

\section{Introduction}

With 450 papers that include the words ``fundamental plane'' in their
titles, our purpose is less to say something new than to reassemble some of
what has been said before and frame it in a way that renders the fundamental
plane somewhat less mysterious.

If one measures stellar velocity dispersions, $\sigma_*$, effective
radii, $r_e$, and effective surface brightnesses, as measured in some
filter, $I_e$, for elliptical galaxies, the observations lie close to a
plane when plotted in the space spanned by $\log \sigma_*$, $\log r_e$
and $\log I_e$ \cite[(Djorgovski \& Davis 1987)]{Djorgovski87}.
Fitting a plane to the data yields three coefficients, as shown in
equation~\ref{fund}.

The effective radius and surface brightness may be combined to produce
a luminosity $L$.  The ellipticals lie along a corresponding plane in
the space spanned by $\log \sigma_*$, $\log r_e$ and $\log L$.  If one
has observations in several filters, one can calculate stellar masses,
$M_*$, 
subject to considerable uncertainty in the initial mass function 
and its trend with velocity dispersion,
and stellar surface densities, $\Sigma_*$.  These also have
associated planes \cite[(Hyde and Bernardi 2009)]{Hyde09}.  For
the present discussion we consider the fundamental plane in the space
spanned by $\log \sigma_*$, $\log r_e$ and $\log M_*$.

\begin{equation}
\matrix{
\underbrace{
a * \log\left(\matrix{
{\rm surface\ brightness\ } I_e\cr 
or\ {\rm luminosity\ } L \cr 
or\ {\rm stellar\ mass\ } M_* \cr 
or\ {\rm surface\ density\ } \Sigma_*
}\right) 
+\  b*\log\left(
{\rm half\ light\ radius\ } r_e
\right)
}_{\rm stellar\ (baryonic)\ matter} \cr
\cr
\underbrace{
+\ c*\log\left(\matrix{
{\rm stellar\ velocity\ dispersion\ } \sigma_*
}\right)}_{{\rm thesis:\ proxy\ for\ dark\ matter\ halo\ dispersion}\ \sigma_{\rm DM}} = 1 }
\label{fund}
\end{equation}

Two of these quantities, the effective radius and the stellar mass,
describe a manifestly baryonic component of elliptical galaxies -- the
stars. We argue here that the third quantity, the stellar velocity
dispersion, is a proxy for the velocity dispersion in the galaxy's
dark matter halo.

\section{The virial theorem}

Astronomers almost always use the virial theorem in the form appropriate
to the global properties of equilibrium systems,
\begin{equation}
2T = -U \quad .
\end{equation}
By contrast, physics texts \cite[(e.g. Goldstein 1980)]{Goldstein80}
often present the virial theorem in a form 
appropriate to the orbit of a single star,
\begin{equation}
\left< v^2\right>_{\rm any\ object\ \ \,} 
= \left<\vec r \cdot \vec \nabla \Phi \right>_{\rm the\ orbit} \quad ,
\end{equation}
where $\left< v^2\right>$ is the twice the orbit averaged kinetic energy
per unit mass, $\vec \nabla \Phi$ is the gradient of the gravitational
potential and $\vec r$ is position.  The orbit averaged kinetic energies
for stars and dark matter particles, are respectively,
\begin{equation}
\left< v^2\right>_{\rm any\ star\ \ \ \ \ } 
\approx \left<\vec r \cdot \vec \nabla \Phi \right>_{10^0 r_e} \quad {\rm and}
\end{equation}
\begin{equation}
\left< v^2\right>_{\rm dark\ particle} 
\approx \left<\vec r \cdot \vec \nabla \Phi \right>_{10^1 r_e} ~\quad\quad .
\end{equation}
The stellar and dark matter
velocity dispersions are the product of the {\it same} gravitational potential
and differ {\it only} insofar as the quantity  
$\left<\vec r \cdot \vec \nabla \Phi \right>$ varies with radius.

But there are several lines of evidence that point to little or no
variation of $\left<\vec r \cdot \vec \nabla \Phi \right>$ with
radius.

\section{The spheroid-halo conspiracy}

It was the observation that spiral galaxies have circular velocities
in excess of what was expected from their observable baryons that
first lead to the conclusion that they were embedded in dark matter
halos.  But beyond that, {\sc H I} rotation curves are remarkably
flat, leading \cite[van Albada and Sancisi (1986)]{vanAlbada86} to
conclude that there was a ``disk-halo conspiracy.''  The baryons
dissipate to the point where they just compensate for what would otherwise
be a decline in the rotation curve in the absence of baryons.  

\cite[Gavazzi et al (2007)]{Gavazzi07} subsequently used a combination
of strong and weak lensing to show that there is likewise
a ``spheroid-halo conspiracy'' for early-type systems.  The gravitational
potentials are isothermal giving  $\left< v^2\right> \sim$ constant.
\cite[Humphrey and Buote (2010)]{Humphrey10} use X-ray observations
and likewise find a conspiracy producing isothermal potentials.
Stellar velocity dispersions are therefore telling us the velocity
dispersion of the dark matter.  

We have marked up the fundamental plane equation~\ref{fund} to 
show that while the effective radius and stellar mass are telling
us about the baryons, the stellar velocity dispersion is telling
us the velocity dispersion of the dark matter halo, $\sigma_{DM}$.

\section{A fundamental {\it line} for dark matter halos}

Dark matter halos are often characterized by their virial radii,
$r_{200}$ and the mass within that radius, $M_{200}$.  It is natural
to ask whether these, along with $\sigma_{DM}$, likewise organize
themselves into a plane.  They do not.  Instead they lie along a tight
line \cite[(Diemer et al 2013)] {Diemer13}.  They are governed by a
single parameter, an overall scale.  Parameterizing the line by the
maximum observed circular velocity, one has 
\begin{equation}
M_{200} \sim V_{max}^3 \quad {\rm and} \quad r_{200} \sim V_{max} \quad.
\end{equation}
The first of these is a consequence of defining halos in terms of densities
while the second is consequence of the ongoing nature of virialization
and the measurement of the virial radius at the same time for all halos.

Now if a stellar velocity dispersion tells us the velocity dispersion
of the dark matter halo, and if that in turn tells us the mass and radius
of the dark matter halo, then we can scale the stellar mass by the
mass in dark matter (or alternatively, the total mass), and we can
likewise scale the effective radius by the dark matter radius,

\section{New axes for the stellar mass fundamental plane}

This gives us new axes for the fundamental plane, as shown in Figure 1.
One axis gives us the overall scale of the system, and a second axis tells
us what fraction of the the baryons that have been incorporated
into stars, on the assumption that the system started out with
the cosmological baryon fraction. 

\begin{figure}[t]
 \centerline{
 \scalebox{0.65}{
 \includegraphics[angle=0]{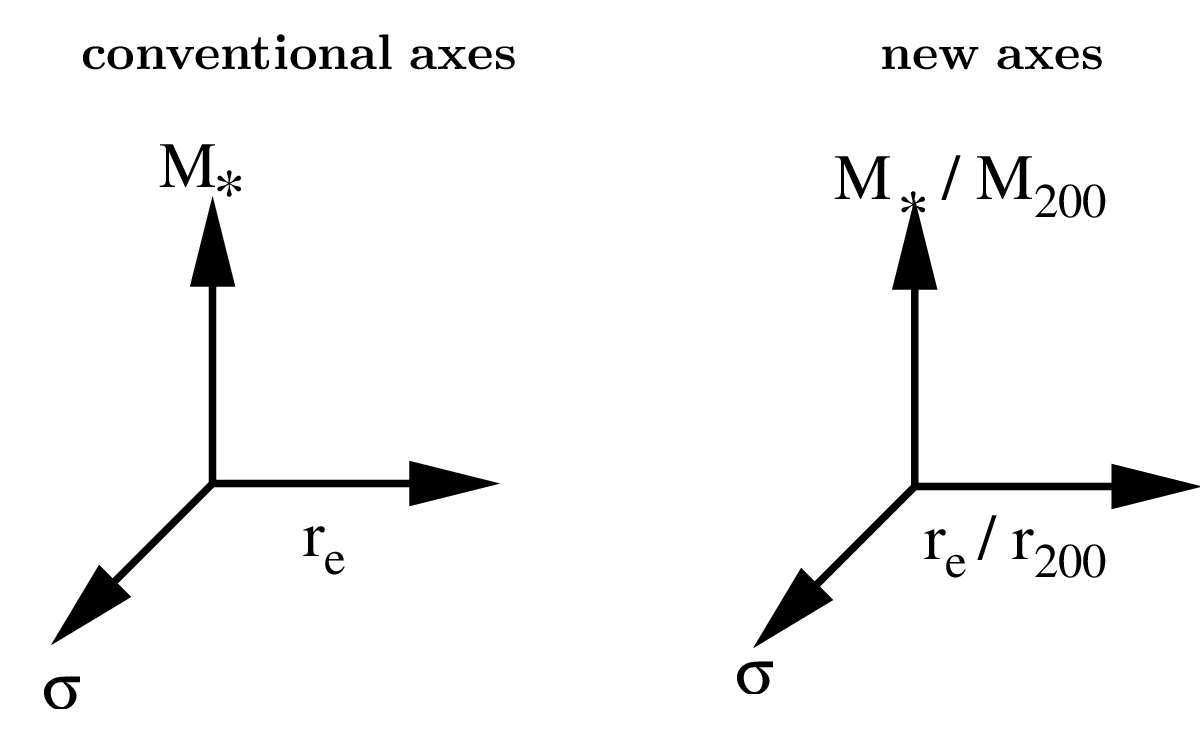}}
 }
  \caption{Axes for the stellar mass fundamental plane,
with dimensioned effective radii and stellar masses
unscaled on the left and 
scaled by the
virial radius and mass of the halo (determined
from the stellar velocity dispersion) on the right.
  }\label{fig:oldvsnew}
\end{figure}

While the above argument invokes the spheroid-halo conspiracy in its
strongest form, with $\sigma_{DM} = \sigma_*$, the argument can 
still be made as long as there is a well defined relation between
stellar velocity dispersion and halo dispersion.

\section{Why not a fundamental line for ellipticals?}

The stellar components of elliptical galaxies are clearly not as
simple as the halos.  While the baryons and dark matter were once
nearly uniformly distributed, some baryons dissipated and formed stars
at the centers of the halos.  The $M_*/M_{200}$ axis tells the fraction of
stars that did so.  The $r_e/r_{200}$ axis is governed both by the dissipation
prior to star formation and by the subsequent remixing of stars by mergers.

If these three processes -- dissipation, star formation and remixing
-- were strictly governed by the scale of the halo, the ellipticals
would lie along a line.  If there were a stochastic component to each
of them, the scatter would thicken the line into a sausage.  The fact
that the ellipticals are confined to a plane indicates that the three
processes are coupled.

\section{Antecedents}

\cite[Zaritsky et al (2008)]{Zaritsky08} anticipated
our representation of the fundamental plane.  They defined an ``efficiency'',
$M_*/M_{\rm baryonic}$, which differs from our choice of coordinate by a
factor of the cosmic baryon fraction.  And they defined a
``concentration'', $r_{200}/r_e$, which is just the inverse of our
coordinate.  They plot both, separately, against stellar velocity
dispersion, but do not examine the three dimensional distribution
and do not argue that the planarity of the elliptical
distribution indicates coupling.  But it is clear from their
discussion that a coupling is both possible and likely.

\section{Two postdictions}

There is considerable evidence that the gravitational potential inside
the effective radius,
and hence the velocity dispersion, is dominated by stars
rather than dark matter
(e.g. \cite[Treu \& Koopmans, 2004]{Treu04}, 
\cite[Mediavilla et al, 2009]{Mediavilla09}).  Measurement of the stellar
velocity dispersion beyond the effective radius might therefore
reflect  the halo dispersion yet more closely than the central 
measurements that are typically used.  

\cite[Schechter et al (2014)]{Schechter14} found that if one used
Einstein ring radii to calculate equivalent velocity dispersions for
the SLACS \cite[(Auger et al 2010)]{Auger10} sample, the resulting
fundamental plane is substantially tighter than the one constructed
using measured stellar velocity dispersions.  It was this result that
led to the present formulation, so it cannot count as confirmation.

Along the same line, \cite[Falc\'on-Barosso and van de Ven
  (2011)]{Falcon11} find that the they get a tighter fundamental plane
when they use velocity dispersions measured out to $r_e$ than the one
they get using dispersions interior to $r_e/8$.  More extensive
samples, and samples extending out to yet larger radii, are on the
horizon.


\begin{thebibliography}{}

\bibitem[Auger et al.(2010)]{Auger10} Auger, M.~W., Treu, T., 
Bolton, A.~S., et al. 2010, \textit{ApJ}, 724, 511 

\bibitem[Diemer et al.(2013)]{Diemer13} Diemer, B., Kravtsov, 
A.~V., \& More, S. 2013, \textit{ApJ}, 779, 159 

\bibitem[Djorgovski \& Davis(1987)]{Djorgovski87} Djorgovski, S.,
\& Davis, M. 1987, \textit{ApJ}, 313, 59 

\bibitem[Falc\'on-Barroso et al.(2011)]{Falcon11} 
Falc\'on-Barroso, J., van de Ven, G., Peletier, R.~F., et al.\ 2011,
\textit{MNRAS}, 417, 1787 

\bibitem[Gavazzi et al.(2007)]{Gavazzi07} Gavazzi, R., Treu, T., 
Rhodes, J.~D., et al. 2007, \textit{ApJ}, 667, 176 

\bibitem[Goldstein]{Goldstein80} Goldstein, H. 1980, \textit{Classical
Mechanics, 2nd ed.} (Addison-Wesley), p. 85

\bibitem[Hyde \& Bernardi(2009)]{Hyde09} Hyde, J.~B.,
\& Bernardi, M. 2009, \textit{MNRAS}, 396, 1171 

\bibitem[Humphrey \& Buote(2010)]{Humphrey10} Humphrey, P.~J., 
\& Buote, D.~A. 2010, \textit{MNRAS}, 403, 2143 

\bibitem[Mediavilla et al.(2009)]{Mediavilla09} Mediavilla, E.,
Mu{\~n}oz, J.~A., Falco, E., et al.\ 2009, \textit{ApJ}, 706, 1451

\bibitem[Schechter et al.(2014)]{Schechter14} Schechter, P.~L., \
Pooley, D., Blackburne, J.~A., \& Wambsganss, J.\textit{ApJ}, 793, 96

\bibitem[Treu \& Koopmans(2004)]{Treu04} Treu, T., 
\& Koopmans, L.~V.~E. 2004, \textit{ApJ}, 611, 739 

\bibitem[van Albada \& Sancisi(1986)]{vanAlbada86} van Albada, T.~S., 
\& Sancisi, R. 1986, \textit{Royal Society of London Philosophical Transactions Series A}, 320, 447 

\bibitem[Zaritsky et al.(2008)]{Zaritsky08} Zaritsky, D., \
Zabludoff, A.~I., \& Gonzalez, A.~H. 2008, \textit{ApJ}, 682, 68 

\end{thebibliography}
\end{document}